\documentclass[12pt,preprint]{aastex}

\shorttitle{Vacuum gap ...} \shortauthors{Gil}

\received{2002 March 26}
\begin{document}
\title{On the formation of inner vacuum gaps in radio pulsars}
\author{Janusz Gil\altaffilmark{1} \& George I. Melikidze\altaffilmark{1}}
\email{jag@astro.ia.uz.zgora.pl} \altaffiltext{1}{Institute of
Astronomy, University of Zielona G\'ora, Lubuska 2, 65-265,
Zielona G\'ora, Poland}

\newcommand{\dmdt}{{\mbox{{\rm M}$_{\odot}$}} {\rm yr}$^{-1}$}
\newcommand{\gcc}{{\rm g} \, {\rm cm}^{-3}}
\newcommand{\rn}{$\rho_{\rm nuc}$}
\newcommand{\msun}{\mbox{{\rm M}$_{\odot}$}}
\newcommand{\mdot}{\mbox{$\dot{M}$}}
\newcommand{\rd}{$\rho_{\rm drip}$}
\def\be{\begin{equation}}
\def\ee{\end{equation}}
\def\lesssim{\raisebox{-0.3ex}{\mbox{$\stackrel{<}{_\sim} \,$}}}
\def\gtrsim{\raisebox{-0.3ex}{\mbox{$\stackrel{>}{_\sim} \,$}}}

\begin{abstract}
The problem of formation of the Ruderman-Sutherland type inner
vacuum gap in neutron stars with ${\bf\Omega}\cdot{\bf B}<0$ is
considered. It is argued by means of the condition $T_i/T_s>1$
(where $T_i$ is the critical temperature above which
$^{56}_{26}$Fe ions will not be bound at the surface and $T_s$ is
the actual temperature of the polar cap surface heated by the
back-flow of relativistic electrons) that the inner vacuum gap can
form, provided that the actual surface magnetic field is extremaly
strong ($B_s\gtrsim 10^{13}$~G) and curved (${\cal R}<10^6$~cm),
irrespective of the value of dipolar component measured from the
pulsar spin down rate. Calculations are carried out for pulsars
with drifting subpulses and/or periodic intensity modulations, in
which the existence of the quasi steady vacuum gap discharging via
${\bf E}\times{\bf B}$ drifting sparks is almost unavoidable.
Using different pair-production mechanisms and different estimates
of the cohesive energies of surface iron ions, we show that it is
easier to form the vacuum gap controlled by the resonant inverse
Compton scaterring seed photons than by the curvature radiation
seed photons.
\end{abstract}

 \section{Introduction}

The consecutive  subpulses in a sequence of single pulses of a
number of pulsars change phase systematically between adjacent
pulses, forming apparent driftbands of the duration from several
to a few tenths of pulse periods. The subpulse intensity is also
systematically modulated along driftbands, typically increasing
towards the pulse centre. In some pulsars, which can be associated
with the central cut of the line-of-sight trajectory, only the
periodic intensity modulation is observed, without any systematic
change of subpulse phase. On the other hand, the clear subpulse
driftbands are found in pulsars associated with grazing
line-of-sight trajectories. These characteristics strongly suggest
an interpretation of this phenomenon as a system of
subpulse-associated beams rotating slowly around the magnetic
axis. \citet[][ hereafter RS75]{rs75} proposed a natural
explanation of the origin of subpulse drift, which involved a
number of isolated ${\bf E}\times{\bf B}$ drifting sparks
discharging the quasi steady vacuum gap formed above the polar cap
of the pulsar with ${\bf\Omega}\cdot{\bf B}<0$, in which
$^{56}_{26}$Fe ions were strongly bound at the surface. Although
the original idea of RS75 associating the rotating sub-beams with
the circulating sparks is still regarded as the best model of
drifting subpulse phenomenon, their vacuum gap was later
demonstrated to suffer from the so-called binding energy problem
\citep[for review see][]{as91,um95,xqz99}. In fact, the cohesive
energies of $^{56}_{26}$Fe ions used by RS75 proved largely
overestimated and the inner vacuum gap envisioned by RS75 was
impossible to form. However, it is worth emphasizing that RS75
considered the canonical surface dipolar magnetic fields with
values determined from the pulsar spindown rate, although they
implicitly assumed small radii of curvature ${\cal R}\sim 10^6$~cm
required by the pair creation conditions, which is inconsistent
with a purely dipolar field. Recently Gil \& Mitra (2001;
hereafter GM01) revisited the binding energy problem an found that
the formation of a vacuum gap is in principle possible, although
it requires an extremely strong non-dipolar surface magnetic field
$B_s=bB_d$, where the coefficient $b \gg 1$ in a typical pulsar,
$B_d=6.4\times 10^{19}(P\dot{P})^{0.5}{\rm G}=2\times
10^{12}(P\dot{P}_{-15})^{0.5}$~G is the dipolar field at the pole,
$P$ is the pulsar period in seconds, $\dot{P}$ is the period
derivative and $\dot{P}_{-15}=\dot{P}/10^{-15}$.

In a superstrong surface magnetic field $B_s>0.1B_q$, where
$B_q=4.414\times 10^{13}$~G, the asymptotic approximation of
\citet{e66} used by RS75 in derivation of the height of quasi
steady vacuum gap is no longer valid. In fact, in such strong
field the high energy $E_f=\hbar\omega$ photons produce
electron-positron pairs at or near the kinematic threshold
$\hbar\omega=2mc^2/\sin\theta$, where $\sin\theta=h/{\cal R}$, $h$
is the gap height, and ${\cal R}={\cal R}_610^6$~cm is the radius
of curvature of surface magnetic field lines \citep[e.g.][]{dh83},
$\hbar$ is the Planck constant, $c$ is the speed of light, $m$ and
$e$ are the electron mass and charge, respectively. The vacuum gap
formed under such conditions was called the Near Threshold Vacuum
Gap (hereafter NTVG) by GM01. They considered two kinds of high
energy seed photons dominating the $e^-e^+$ pair production: the
Curvature Radiation (CR) photons with energy
$\hbar\omega=(3/2)\hbar\gamma^3c/{\cal R}$ \citep[RS75;][]{zq96},
and resonant Inverse Compton Scattering (ICS) photons with energy
$\hbar\omega=2\gamma\hbar eB_s/mc$ \citep{zq96,zetal97}, where
$\gamma$ is a typical Lorentz factor of particles within the gap.
The corresponding vacuum gap is called the Curvature Radiation
dominated (CR-NTVG) and the Inverse Compton Scattering dominated
(ICS-NTVG), respectively. GM01 estimated the characteristic
heights of both CR-NTVG and ICS-NTVG. In this paper we further
refine these estimates by including the general relativistic (GR)
effects of inertial frame dragging (IFD) and considering the heat
flow conditions within the thin uppermost surface layer of the
polar cap. Moreover, we use a broader range of cohesive energies
of surface $^{56}_{26}$Fe ions. The obtained VG models are applied
to pulsars with drifting subpulses and/or periodic intensity
modulations, in which the presence of ${\bf E}\times {\bf B}$
drifting spark discharges seems almost unavoidable (Deshpande \&
Rankin 1999, 2001; Vivekanand \& Joshi 1999, and Gil \& Sendyk
2000).

\section{Near threshold vacuum gap formation}

If the cohesive energy of $^{56}_{26}$Fe ions is large enough to
prevent them from thermionic (this section) or field emission
(Appendix A), a vacuum gap forms right above the polar cap with a
characteristic radius $r_p=b^{-0.5}10^4P^{-0.5}$~cm
\citep[e.g.][]{gs00}. \citet{jlk01} pointed out that GR effect of
IFD \citep{mt92,mh97} should affect the RS75 type models with a
vacuum gap above the polar cap. Although \citet{jlk01} did not
investigate the problem, they implicitly suggested that the
electric fields distorted by GR effects make formation of
``starved'' inner magnetospheric regions even more difficult than
in the flat space case. However, \citet{zhm00} demonstrated that
although GR-IFD effect is small, it nevertheless slightly helps
formation of VG above the polar caps. In other words, GR modified
potential drop within VG is slightly lower than in the flat space
case. Below we confirm this finding for NTVG conditions, with a
very strong and complicated surface magnetic field.

The gap electric field ${\bf E}_\|$ along ${\bf B}_s$ results from
a deviation of the local charge density $\rho\approx 0$ from the
corotational charge density $\rho_c=(\zeta/\alpha)\rho_{GJ}$,
where $\rho_{GJ}=-{\bf\Omega}\cdot{\bf B}_s/(2\pi c)$ is the flat
space-time Goldreich-Julian (1969) charge density,
$\Omega=2\pi/P$, $\zeta=1-\kappa_g$,
$\kappa_g\sim(r_g/R)(I/MR^2)$, $\alpha=(1-r_g/R)^{1/2}$ is the
redshift factor, $r_g$ is the gravitational radius, $M$ is the
neutron star mass and $I$ is the neutron star moment of inertia.
The potential $V$ and electric fields $E_\|$ within a gap are
determined by GR-analog \citep{mt92} of the one dimensional the
Poisson equation $(1/\alpha)d^2V/dz^2=-4\pi(\rho-\rho_c)=2\Omega
B_s\zeta/(c\alpha)$, with a boundary condition $E_\|(z=h)=0$,
where $h$ is the height of an infinitesimal gap. The solution of
the Poisson equation gives \be E_\|(z)=\zeta(2\Omega
\frac{B_s}{c})(h-z) ,\ee and \be \Delta V=\zeta\frac{\Omega
B_s}{c}h^2 .\ee In further calculations we will adopt a typical
value of the correction factor $\zeta\sim 0.85$ (corresponding to
$M=1M_\odot$, $R=10^6$~cm and $I=10^{45}{\rm g\ cm}^2$), although
its value can be as low as 0.73 \citep{zhm00}. Thus, the potential
drop within the actual gap can be 15 to 27 percent lower than in
the conventional RS75 model.

The polar cap surface of the pulsar with ${\bf\Omega}\cdot{\bf
B}<0$ is heated by a back-flow of relativistic electrons
(accelerated in the parallel electric field $E_\|$) to the
temperature $T_s=k^{1/4}(e\Delta V\dot{N}/\sigma\pi r_p^2)^{1/4}$,
where $\Delta V$ is described by equation~(2), $\dot{N}=\pi
r_p^2B_s/eP$ is the Goldreich-Julian kinematic flux and the heat
flow coefficient $0.1\lesssim k<1$ is described in Appendix B. The
thermal condition for the vacuum gap formation can be written in
the form $T_i/T_s>1$, where $T_s$ is the actual surface
temperature described above, and $T_i=\Delta\varepsilon_c/30$k is
the iron critical temperature above which $^{56}_{26}$Fe ions are
not bound on the surface (that is, a copious amounts of
$^{56}_{26}$Fe ions at about Goldreich-Julian density are
available for thermionic ejection from the surface), where
$\Delta\varepsilon_c$ is the cohesive energy of condensed
$^{56}_{26}$Fe matter in the neutron star surface and
k$=1.38\times 10^{-23}{\rm JK}^{-1}$ is the Boltzman constant
\citep{cr80,um95}. The properties of condensed matter in very
strong magnetic fields characteristic of a neutron star surface
have been investigated by many authors using different examination
methods \citep[for review see][]{um95}. There exist many
discrepancies in determination of the cohesive energy
$\Delta\varepsilon_c$ and in this paper we refer to the two
papers, representing the limiting extreme cases. Abrahams \&
Shapiro (1991; AS91 henceforth) estimated
$\Delta\varepsilon_c=0.91$~keV, 2.9 keV and 4.9 keV for
$B_s=10^{12}$~G, $5\times 10^{13}$~G and $10^{13}$~G,
respectively. These values were approximated by \citet{um95} in
the form $\Delta\varepsilon_c\simeq(0.9{\rm keV})(B_s/10^{12}~{\rm
G})^{0.73}$, which leads to critical temperatures \be
T_i=(3.5\times 10^5{\rm K})(B_s/10^{12}{\rm G})^{0.73}=(6\times
10^5)b^{0.73}(P\dot{P}_{-15})^{0.36}~{\rm K} .\ee On the other
hand, Jones (1986; J86 henceforth) obtained much lower cohesive
energies $\Delta\varepsilon_c=0.29$~keV, 0.60 keV and 0.92 keV for
$B_s=2\times 10^{12}$~G, $5\times 10^{12}$~G and $10^{13}$~G,
respectively. They can be approximated by
$\Delta\varepsilon_c\simeq(0.18\ {\rm keV})(B_s/10^{12}{\rm
G})^{0.7}$ and converted to critical temperatures  \be
T_i=(0.7\times 10^5{\rm K})(B_s/10^{12}{\rm G})^{0.7}=(1.2\times
10^5)b^{0.7}(P\cdot\dot{P}_{-15})^{0.36}~{\rm K}. \ee Below we
consider the condition $T_i/T_s>1$, using both expressions for
critical temperatures described by equations~(3) and (4), for CR-
and ICS-dominated NTVG models, separately.

\subsection{CR-NTVG}

In this case the gap height $h=h_{CR}$ is determined by the
condition that $h=l_{ph}$, where $l_{ph}\approx\sin\theta\ {\cal
R}=(B_\perp/B_s){\cal R}$ is the mean free path for pair
production by a photon propagating at an angle $\theta$ to the
local surface magnetic field (RS75). The CR-NTVG model is
described by the following parameters: the height of a quasi
steady gap \be h_{CR}=(3\times 10^3)\zeta^{-3/7}{\cal
R}_6^{2/7}b^{-3/7}P^{3/14}\dot{P}_{-15}^{-3/14}~{\rm cm} ,\ee
(notice typographical errors in eq.~[6] of GM01, which are
corrected here), the gap potential drop \be \Delta
V_{CR}=(1.2\times 10^{12})\zeta^{1/7}{\cal
R}_6^{4/7}b^{1/7}P^{-1/14}\dot{P}_{-15}^{1/14}\ {\rm V} ,\ee and
the surface temperature \be T_s=(3.4\times
10^6)\zeta^{1/28}k^{1/4}{\cal
R}_6^{1/7}b^{2/7}P^{-1/7}\dot{P}_{-15}^{1/7} ~{\rm K}. \ee The
thermal condition $T_i/T_s>1$ for the formation of CR-NTVG leads
to a family of critical lines on the $P-\dot{P}$ diagram (see
Fig.~1 in GM01) \be \dot{P}_{-15}\geq {\cal
A}^2\zeta^{0.16}k^{1.14}{\cal R}_6^{0.64}b^{-2}P^{-2.3} ,\ee where
${\cal A}=(2.7\times 10^3)^{1/2}=52$ for AS91 case (eq.~[3]) and
${\cal A}=(3.96\times 10^6)^{1/2}=1990$ for J86 case (eq.~[4]).
Alternatively, one can find a minimum required surface magnetic
field $B_s=bB_d$ expressed by the coefficient $b$ in the form \be
b_{min}^{CR}={\cal A} \zeta ^{0.08}k^{0.57}{\cal
R}_6^{0.32}P^{-1.15}\dot{P}_{-15}^{-0.5} .\ee

\subsection{ICS-NTVG}

In this case the gap height $h=h_{ICS}$ is determined by the
condition $h=l_{ph}\sim l_e$, where $l_e$ is the mean free path of
the electron to emit a photon with energy
$\hbar\omega=2\gamma\hbar eB_s/mc$ \citep[][ GM01]{zhm00}. The
ICS-NTVG model is described by the following parameters: the
height of a quasi steady gap \be h_{ICS}=(5\times
10^3)\zeta^{0.14}k^{-0.07}{\cal
R}_6^{0.57}b^{-1}P^{-0.36}\dot{P}_{-15}^{-0.5}~{\rm cm},\ee the
gap potential drop \be \Delta V_{ICS}=(5.2\times 10^{12})
\zeta^{0.72}k^{-0.14}{\cal
R}_6^{1.14}b^{-1}P^{-1.22}\dot{P}_{-15}^{-0.5}~{\rm V},\ee and the
surface temperature \be T_s=(4\times
10^6)\zeta^{0.18}k^{0.21}{\cal R}_6^{0.28}P^{-0.43} ~{\rm K}. \ee
The thermal condition $T_i/T_s>1$ for the formation of ICS-NTVG
leads to a family of critical lines on the $P-\dot{P}$ diagram
(see Fig.~1 in GM01) \be \dot{P}_{-15}\geq {\cal
B}^2\zeta^{0.5}k^{0.7}{\cal R}_6^{0.8}b^{-2}P^{-2.2} ,\ee where
${\cal B}=(2\times 10^2)^{1/2}=14$ for AS91 case (eq.~[3]) and
${\cal B}=(1.69\times 10^4)^{1/2}=130$ for J86 case (eq.~[4]).
Alternatively, one can find a minimum required surface magnetic
field $B_s=bB_d$ expressed by the coefficient $b$ in the form \be
b_{min}^{ICS}={\cal B}\zeta^{0.25}k^{0.34}{\cal
R}_6^{0.39}P^{-1.1}\dot{P}_{-15}^{-0.5} .\ee
A comparison of spark developing time scales in ICS- and CR-dominated
vacuum gaps is presented in Appendix C.

\subsection{NTVG development in pulsars with drifting subpulses}

GM01 examined the $P-\dot{P}$ diagram (their Fig.~1) with 538
pulsars from the Pulsar Catalog \citep{tetal93} with respect to
the possibility of VG formation by means of the condition
$T_i/T_s$, with $T_i$ determined according to AS91 (eq.~[3]). They
concluded that formation of VG is in principle possible, although
it requires a very strong and curved surface magnetic field
$B_s=b\cdot B_d\gtrsim 10^{13}$~G, irrespective of the value of
$B_d$. Here we reexamine this problem by means of equations~(9)
and (14), with critical temperatures corresponding to both AS91
case (${\cal A}=52$ and ${\cal B}=14$) and J86 case (${\cal
A}=1990$ and ${\cal B}=130$), for CR and ICS seed photons,
respectively. Therefore, we cover practically almost the whole
range of critical temperatures between the two limiting cases,
which differ by a factor of five between each other. We adopt the
GR-IFD correction factor $\zeta=0.85$ \citep{hm98,zhm00}, the
normalized radius of curvature of surface magnetic field lines
$0.01\leq{\cal R}_6\leq 1.0$ (GM01 and references therein) and the
heat flow coefficient $0.1\leq k<1.0$ (Appendix B).

The results of calculations of NTVG models for 42 pulsars with
drifting subpulses and/or periodic intensity modulations (after
Rankin 1986) are presented in Fig.~1. We calculated the ratio
$B_s/B_q=0.0453\ b(P\dot{P}_{-15})^{0.5}$, where $B_q=4.414\times
10^{13}$~G and the coefficient $b$ is determined by equation~(9)
and equation~(14) for CR and ICS seed photons, respectively.
Noting that the functional dependence on $P$ and $\dot{P}_{-15}$
in both equations is almost identical, we sorted the pulsar names
(shown on the horizontal axes) according to the increasing value
of the ratio $B_s/B_q$ (shown on the vertical axes). Calculations
were carried out for $B_s/B_q\geq 0.1$, since below this value the
near-threshold treatment is no longer relevant. On the other hand,
the values of $B_s/B_q$ are limited by the photon splitting
threshold, which is roughly about $10^{14}$~G in typical pulsars
\citep[][ see also astro-ph/0102097]{bh98,bh01,z01}. Therefore,
the physically plausible NTVG models lie within the shaded areas,
with the upper boundary determined by the photon-splitting level
and the lower boundary determined by the lowest ICS line
calculated for $k=0.1$ (see below). The left hand side of Fig.~1
corresponds to AS91 case (${\cal A}=52$ and ${\cal B}=14$ in
eqs.~[9] and [14], respectively) and the right hand side of Fig.~1
corresponds to J86 case (${\cal A}=1930$ and ${\cal B}=130$ in
eqs.~[9] and [14], respectively). Four panels in each side of the
figure correspond to different values of the radius of curvature
${\cal R}_6=1.0$, 0.1. 0.05 and 0.01 from top to bottom,
respectively (indicated in the upper corner of each panel). Two
sets of curved lines in each panel correspond to the CR (thin
upper lines) and ICS (thick lower lines) seed photons,
respectively. Three different lines within each set correspond to
different values of the heat flow coefficient $k=1.0$ (dotted),
$k=0.6$ (dashed) and $k=0.2$ (long dashed), and lower boundary of
shaded areas corresponds to $k=0.1$.

A visual inspection of the model curves within shaded areas in
Fig.~1 shows that in AS91 case ICS-NTVG is favored for larger
radii of curvature ${\cal R}_6>0.05$, while CR-NTVG requires lower
values of ${\cal R}_6<0.1$. In J86 case, only ICS-NTVG
corresponding to ${\cal R}_6<0.1$ can develop. If the actual
cohesive energies correspond to some intermediate case between
AS91 and J86 cases, they will also be associated with ICS-NTVG.
Therefore, we can generally conclude that ICS-NTVG model is
apparently favored in pulsars with drifting subpulses. CR-NTVG is
also possible, although it requires a relatively low cohesive
energies, as well as  an extremely strong ($B_s/B_q\sim 1$) and/or
curved (${\cal R}_6\sim 0.01$) surface magnetic field at the polar
cap.

\section{Conclusions}

There is a growing evidence that the radio emission of pulsars
with systematically drifting subpulses (grazing cuts of the
line-of-sight) or periodic intensity modulations (central cuts of
the line-of-sight) is based on the inner vacuum gap developed just
above the polar cap \citep{dr99,dr01,vj99,gs00}. To overcome the
binding energy problem \citet{xqz99,xzq01} put forward an
attractive but exotic conjecture that pulsars showing the drifting
subpulses represent bare polar cap strange stars (BPCSS) rather
than neutron stars. However, as demonstrated in this paper,
invoking the BPCSS conjecture is not necessary to explain the
drifting subpulse phenomenon. The quasi steady vacuum gap, with
either curvature radiation or inverse Compton scattering seed
photons, can form in pulsars with ${\bf\Omega}\cdot{\bf B}<0$,
provided that the actual surface magnetic field at the polar cap
is extremely strong $B_s\sim 10^{13}$~G and curved ${\cal
R}<10^6$~cm, irrespective of the value of dipolar component
measured from the pulsar spindown rate. We have used two sets of
the cohesive (bounding) energies of the surface iron ions: higher
values obtained by AS91 and about five times lower values obtained
by J86. If the actual cohesive energies are close to J86 values,
then only ICS controlled VG can form and CR controlled VG never
forms even under the most extreme conditions (see Fig. 1).

It is worth noting that ICS discussed in this paper is the
so-called "resonant" scattering. The choice of this mode of ICS is
justified by the superstrong magnetic fields relevant for the
near-threshold regime. However, it is clear that the thermal
photons also contribute some pair-producing gamma-rays
\citep[e.g.][]{zetal97}. This process cannot dominate the
breakdown of the gap, since it requires higher surface
temperatures, which may inhibit the formation of VG in the first
place. One can only speculate that the dominating thermal ICS mode
is associated with the pulse nulling phenomenon. These issues
require further investigation.

The pulsars with drifting subpulses and/or periodic intensity
modulation do not seem to occupy any particular region of the
$P-\dot{P}$ diagram (see Fig.~1 in GM01). Rather, they are spread
uniformly all over the $P-\dot{P}$ space, at least for typical
pulsars (excluding young and millisecond pulsars, in which
observations of single pulses are difficult in the first place).
Therefore, it seems tempting to propose that radio emission of all
pulsars should be driven by vacuum gap activities. An attractive
property of such proposition is that the nonstationary sparking
discharges induce the two-stream instabilities that develop at
relatively low altitudes \citep{u87,am98} where the pulsars radio
emission is expected to originate \citep{c78,kg97,kg98}. It is
generally believed that the high frequency plasma waves generated
by the two-stream instabilities can be converted into coherent
electromagnetic radiation at pulsar radio wavelengths
\citep[e.g.][]{mgp00}. With such scenario, all radio pulsars would
require a strong, non-dipolar surface magnetic field at their
polar caps (e.g. Gil et al. 2002 a, b).

\acknowledgments

This paper is supported in part by the KBN Grant 2~P03D~008~19 of
the Polish State Committee for Scientific Research. We are
indebted to anonymous referee for insightful comments and
constructive criticism, which greatly helped to improve the final
version of the paper. We are grateful to Prof. V. Usov and Dr. B.
Zhang for helpful discussions. We also thank E. Gil, U.
Maciejewska and Dr. M. Sendyk for technical help.

\begin{appendix}

\section{Field emission}

The vacuum gap can form in pulsars with ${\bf\Omega}\cdot{\bf
B}<0$ if the actual surface temperature $T_s$ is not high enough
to liberate $^{56}_{26}$Fe ions from the polar cap surface by
means of thermal emission. Now we examine the field (cold cathode)
emission, which is possibly important when thermionic emission is
negligible \citep[e.g.][]{zh00}. The maximum electric field (along
${\bf B}_s$) at the NS surface
$E_\|(max)=\zeta(4\pi/cP)B_sh_{ICS}=(1.25\times
10^9)\zeta^{0.86}b^{-1}{\cal R}_6^{0.57}P^{0.14}~{\rm V}/{\rm
cm}$, or taking into account that $\zeta=0.85$ and $b\gtrsim
2(P\cdot\dot{P}_{-15})^{-0.5}$, $E_\|(max)\leq(5\times 10^8){\cal
R}_6^{0.57}P^{0.64}\dot{P}_{-15}^{0.5}$~V/cm. The critical
electric field needed to pull $^{56}_{26}$Fe ions from the NS
surface is $E_\|(crit)=(8\times
10^{12})(\Delta\varepsilon_c/26~{\rm keV})^{3/2}$~V/cm
\citep{as91,um95,um96}, where the cohesive (binding) energy of
iron ions in a strong surface magnetic field $B_s\sim 10^{13}$~G
is $\Delta\varepsilon_c=4.85$~keV \citep{as91}. Thus,
$E_\|(crit)=6.4\times 10^{11}~{\rm V/cm}>>E_\|\sim 10^9$~V/cm and
no field emission occurs. It is possible that the cohesive energy
is largely overestimated and $\Delta\varepsilon_c$ can be much
smaller than about 4 keV even at $B_s>10^{13}$~G. We can thus ask
about the minimum value of $\Delta\varepsilon_c$ at which the
field emission is still negligible. By direct comparison of
$E_\|(max)$ and $E_\|(crit)$ we obtain
$\Delta\varepsilon_c>40x^{0.67}$~eV, where $x={\cal
R}_6^{0.57}P^{0.64}\dot{P}_{-15}^{0.5}$ is of the order of unity.
Therefore, contrary to the conclusion of \citet{jlk01}, no field
emission is expected under any circumstances.

\section{Heat flow conditions at the polar cap surface}

Let us consider the heat flow conditions within the uppermost
surface layer of the pulsar polar cap above which NTVG can
develop. The basic heat flow equation is \citep[e.g.][]{ec89,u01}
\be C\frac{\partial T}{\partial t}=\frac{\partial}{\partial
x}\left(\kappa_\|\frac{\partial T}{\partial x}\right) ,\ee  where
$C$ is the heat capacity (per unit volume) and
$\kappa_\|>>\kappa_\perp\sim 0$ is the thermal conductivity along
($\|$) surface magnetic field lines (which are assumed to be
perpendicular ($\perp$) to the polar cap surface). The heating of
the surface layer of thickness $\Delta x$ is due to sparking
avalanche with a characteristic development time scale $\Delta t$.
We can write approximately $\partial T/\partial t\approx T/\Delta
t$, $\partial T/\partial x\approx T/\Delta x$, $\partial/\partial
x(\partial T/\partial x)\approx T/\Delta x^2$ and thus $C/\Delta
t\approx \kappa_\|\Delta x^2$. Therefore, the crust thickness
$\Delta x$ heated during $\Delta t$ is approximately $\Delta x
\approx\left({\kappa_\|}\Delta t/{C}\right)^{1/2}$ (within an
uncertainty factor $\delta x$ of 2 or so), where the time scale
corresponds to the spark development time $\Delta t\approx
10\mu{\rm s}$ (RS75). The energy balance equation is
$Q_{heat}=Q_{rad}+Q_{cond}$, where $Q_{heat}=en_{GJ}\Delta
V_{max}$ (e.g. RS75), $Q_{rad}=\sigma T_s^4$ ($T_s$ is the actual
surface temperature and $\sigma=5.67\times 10^{-5}~{\rm erg\
cm}^{-2}K^{-4}s^{-1}$), and $Q_{cond}=-\kappa_\|\partial
T/\partial x\approx -\kappa_\|T_s/\Delta x$. We can now define the
heat flow coefficient $k=Q_{rad}/Q_{heat}<1$, which describes
deviations from the black-body conditions on the surface of the
sparking polar cap. In other words, the value of $k$ describes the
amount of heat conducted beneath the polar cap which cannot be
transferred back to the surface from the penetration depth $\Delta
x$ during the time-scale $\Delta t$ (see Appendix C). The
coefficient $k$ can be written in the form
 \be
k=\frac{1}{1+\kappa_\|/(\sigma T_s^3\Delta
x)}=\frac{1}{1+(\kappa_\|C)^{1/2}/(\sigma T_s^3\Delta t^{1/2})} ,
\ee whose value can be estimated once the parameters $C$,
$\kappa_\|$ and $T_s$ as well as $\Delta x$ or $\Delta t$ are
known.

The matter density at the neutron star surface composed mainly of
$^{56}_{26}$Fe ions \citep[e.g.][]{um95} is $\rho(B_s)\simeq
4\times 10^3(B_s/10^{12}~{\rm G})^{6/5}{\rm g\ cm}^{-3}$
\citep{fetal77}. Thus for $B_s\sim (1\div 3)\times 10^{13}$~G we
have $\rho\sim (0.6\div 2.4)\times 10^5{\rm g\ cm}^{-3}$. The
thermal energy density of $^{56}_{26}$Fe is $U_{Fe}\simeq
2.2\times 10^{19}\rho_5T_8\ {\rm erg\ cm}^{-3}$ \citep{ec89},
where $\rho_5=\rho/(10^5 {\rm g\ cm}^{-3})$ and
$T_8=T_s/(10^8~{\rm K}$). Thus, the heat capacity
$C=dU_{Fe}/dt=10^{-8}dE_{Fe}/dT_8=(1\div 5)\times 10^{11}\rho_5\
{\rm erg\ cm}^{-3} {\rm K}^{-1}$. For $\rho_5\approx 1$ we have
$C\approx 2\times 10^{11}{\rm erg\ cm}^{-3}{\rm K}^{-1}$. The
longitudinal thermal conductivity can be estimated as
$\kappa_\|\approx (2\div 4)\times 10^{13}\ {\rm erg\ s}^{-1}{\rm
cm}^{-1}{\rm K}^{-1}$ \citep[see Fig.~5 in][]{p99}. Thus
$\kappa_\|/C\approx 1.5\times 10^2\ {\rm cm}^2$ and the
penetration depth $\Delta x\approx 10\Delta t^{1/2}s^{-1/2}$~cm.
Since the characteristic spark development time scale $\Delta
t\sim 10 \mu s =10^{-5}$~s (see Appendix C), then $\Delta x\approx
0.03$ ~cm. More generally, the penetration depth can be written as
\be \Delta x=0.03\,\delta
x\left(\frac{\kappa_{13}}{C_{11}}\right)^{1/2} ,\ee where
$\kappa_{13}=\kappa_\|/10^{13}\approx 2\div 4$ and
$C_{11}=C/10^{11}\approx 1\div 5$ and the uncertainty factor
$\delta x\approx 0.5\div 2$. Now the heat flow coefficient can be
written as \be k=\frac{1}{1+5.6\Delta/T_6^3} ,\ee where
$T_6=T_s/10^6$ and $\Delta=(\kappa_{13}C_{11})^{1/2}/\delta
x\approx 0.7\div 6.3$. Figure 2 shows variations of the heat flow
coefficient $k$ versus the surface temperature $T_6$ (in $10^6$~K)
for three values of $\Delta=1$ (upper curve), $\Delta=3$ (middle
curve) and $\Delta=6$ (lower curve). As one can see, for realistic
surface temperatures of a few times $10^6$~K, the values of the
heat flow coefficient $k$ are in the range $0.2\div 0.9$.
\citet{gmm02a} found from independent considerations that in PSR
B0943+10 the heat flow coefficient $k<0.8$, in consistency with
the results obtained in this paper.

The above treatment of the heat transportation corresponds to one
spark event or a relatively short sequence of sparks reappearing
at the same place of the polar cap (modulo the slow ${\bf E}
\times {\bf B}$ drift). If a longer sequence of consecutive sparks
can operate at the same place, deep-layer photons can also diffuse
out of the surface, although after a considerably longer time
\citep{ec89}. The contribution from this component would make $k$
closer to unity. Even if this is the case, ICS driven VG is still
possible to form for all pulsars from our sample, although an
extremely strong and curved surface magnetic field would be
required (the right hand-side panels of Fig. 1).

\section{Characteristic spark development time scale}

The characteristic spark development time scale is defined as the
time interval after which the density of electron-positron spark
plasma grows from essentially zero to the Goldreich-Julian (1969)
value. In the case of CR-induced sparks this time scale is
$\tau_{CR}=(30-40)h_{CR}/c\sim 10\mu s$. The time structure of the
ICS-induced spark avalanche may be different. This follows from
the fact that in the CR-induced avalanche the mean free path $l_e$
of electron/positron to generate one pair-producing gamma photon
is much smaller than the mean free path $l_{ph}$ of gamma photon
to produce an electron-positron pair, while in the ICS-induced
avalanche $l_e$ is comparable with $l_{ph}$ \citep{zhm00}. Let us
introduce two time scales: the time $t_p=l_e/c$ for a charged
particle (electron or positron) to emit a gamma-quantum, and the
time $t_q=l_{ph}/c$ for a gamma-quantum to create an
electron-positron pair. In both CR- and ICS-induced gap discharges
the condition $t_p\lesssim t_q$ holds ($t_p\ll t_q$ in the CR case
and $t_p\sim t_q$ in the ICS case), which suggests that the
avalanche time scale is determined by $t_q$. In fact, one can
assume that the cascade consists of a sequence of $n(t)\approx
t/t_q$ generations, occurring between the initial time $t=0$ and
the current time $t$. Thus, the number of particles created during
the n-th generation is
$N(t)=(2t_q/t_p)^{n(t)}=(2t_q/t_p)^{t/t_q}$. This estimate
represents the following scheme of generation-by-generation
cascade development: $(2t_q/t_p)\rightarrow(2t_q/t_p)^2\rightarrow
...\rightarrow(2t_q/t_p)^{t/t_q}$. For the ICS-induced cascade we
have $t_p\approx t_q\approx h_{ICS}/c$, while for the CR-induced
case $t_q=h_{CR}/c$ and $t_p\approx 2\pi\nu_m\hbar/L_{CR}\approx
3\times 10^{-3}{\cal R}_6/\gamma$, where
$L_{CR}=(2/3)ce^2\gamma^4/{\cal R}$ is the power of the curvature
radiation at the characteristic frequency
$\nu_m=0.4\gamma^3c/R\approx 2\times 10^9\gamma^3/{\cal R}s^{-1}$,
$\gamma=e\Delta V/mc^2=10^7\zeta^{0.14}{\cal R}_6^{0.57}b^{0.14}$
is the Lorentz factor of relativistic charged particles, $\Delta
V=\Delta V_{CR}$ is the NTVG potential drop described by eq.(6),
and ${\cal R}={\cal R}_610^6$~cm is the radius of curvature of the
surface magnetic field $B_s=bB_d$ (we assumed $P=\dot{P}_{-15}=1$
for simplicity). Here $e$ is the electron charge, $c$ is the speed
of light and $\hbar $ is the Planck constant.

The total number of particles created during the time interval $t$
is $N=\int_0^x{N(x)dx}$, where $x=t/t_q$. Thus, for the CR-induced
case in which $t_q/t_p\approx h_{CR}/(t_pc)$ \be N_{{CR}}\sim
\frac{\left( 2\frac{h_{{CR}}}{t_{p}c}\right)
^{\frac{tc}{h_{{CR}}}}}{\ln \left( 2\frac{h_{{CR}}}{t_{p}c}\right)
}, \ee and for the ICS-induced case in which $t_q/t_p\approx 1$
\be N_{{ICS}}\sim \frac{\left( 2\right)
^{\frac{tc}{h_{{ICS}}}}}{\ln 2}, \ee where $h_{CR}$ and $h_{ICS}$
are described in eqs.(5) and (10), respectively.

The cascade terminates at the time $t=\tau$ when the spark plasma
density reaches the Goldreich-Julian density $n_{GJ}$, that is $
N_{ICS}/h_{ICS}^3=N_{CR}/h_{CR}^3 \approx n_{GJ}$. Using this
condition and eqs.(C1) and (C2) we obtain the ICS-induced spark
development time scale $\tau_{ICS}$ as a function of CR-induced
spark development time scale $\tau_{CR}$ in the form \be
\tau_{{ICS}}= \frac{h_{{ICS}}}{h_{{CR}}\ln2}\ln \left(
2\frac{h_{{CR}}}{t_{p}c}\right)
\tau_{{CR}}+\frac{h_{{ICS}}}{c\ln2}\ln \left( \frac{\ln 2}{\ln
\left( 2\frac{h_{{CR}}}{t_{p}c}\right) }\left(
\frac{h_{{ICS}}}{h_{{CR}}}\right) ^{3}\right).\label{tics} \ee
This time scale depends also on other pulsar parameters, mostly on
$b$ and ${\cal R}_6$ (and weakly on $P, \dot{P}, k$ and $\zeta$).
The second term in eq.~(\ref{tics}) is always negative and much
smaller than the first term (for $\tau_{CR}>20h_{CR}/c$; e.g.
RS75). Therefore we obtain an upper limit for the ratio \be
\frac{\tau_{{ICS}}}{\tau_{{CR}}}=2.4\zeta ^{0.57}k^{0.07}{\cal
R}_{6}^{0.28}b^{-0.79}\left( \ln \frac{6.67\times 10^{2}}{\zeta
_{6}^{0.29}{\cal R}_{6}^{0.14}b^{0.07}}\right). \label{tauics}\ee
Fig.~3 shows the upper limits (equality in eq.~[\ref{tauics}]) of
the ratio $\tau_{ICS}/\tau_{CR}$ as a function of
$B_s/B_q=0.0453b$ (see section 2.3 and Fig. 1 for explanation),
for different values of ${\cal R}_6$ and $k$ (for the clarity of
presentation we adopted $P=\dot{P}_{-15}=1$ and $\zeta=0.85$). As
one can see from this figure, the characteristic spark development
time scales of CR- and ICS- induced discharges are comparable.
This is a consequence of the fact that the stronger the magnetic
field the smaller the $l_e$ for ICS, thus the larger the
pair-production multiplicity and the smaller the spark development
time scale $\tau_{{ICS}}$. We can thus generally state that in the
near-threshold regime ICS-driven spark avalanche looks very
similar to the conventional RS75 CR-driven avalanche.

\end{appendix}

\clearpage
\begin{center}
Figure Captions
\end{center}

\figcaption{Models of NTVG for 42 pulsar with drifting subpulses
and/or periodic intensity modulations for AS91 and J86 cohesive
energy calculations. Both curvature radiation (CR) and resonant
inverse Compton radiation (ICS) seed photons generating
electron-positron pairs in a superstrong surface magnetic field
${\bf B}_s$ are considered. The ratio of $B_s/B_q\geq 0.1$ (near
threshold treatment ), where $B_s$ is the actual surface magnetic
field and $B_q=4.4\times 10^{13}$~G is the quantum critical
magnetic field, is presented on the vertical axis. The horizontal
axis corresponds to 42 pulsars with drifting subpulses and/or
periodic intensity modulations. Three different lines correspond
to different values of the heat flow coefficient $k=1.0$ (dotted),
$k=0.6$ (dashed) and $k=0.2)$ (long dashed). Four different panels
correspond to different values of curvature radii from ${\cal
R}_6=1.0$ (top) to ${\cal R}_6=0.01$ (bottom). The physically
plausible models correspond to shaded areas limited by the photon
splitting threshold $B_s\sim 10^{14}$~ G from above and by the
lowest ICS curve (calculated for $k=0.1$) from below.\label{fig1}
}

\figcaption{Heat flow coefficient $k$ versus surface temperature
$T_6=T_s/10^6$~K (see text for explanations). \label{fig2} }

\figcaption{Upper limits for the ratio of development time scales
in ICS- and CR- induced spark discharges of NTVG, as a function of
the surface magnetic field $B_s/B_q$, ranging from $0.1$ (near
threshold treatment) to $2.25$ (photon splitting limit). Four 
values of ${\cal R}_6$=1.0, 0.1, 0.05 and 0.01 (from the top to the bottom)
and two values of $k$=1.0 (solid line) and 0.2 (dashed line) were used.
\label{fig3} }
%\clearpage

\newpage
\plotone{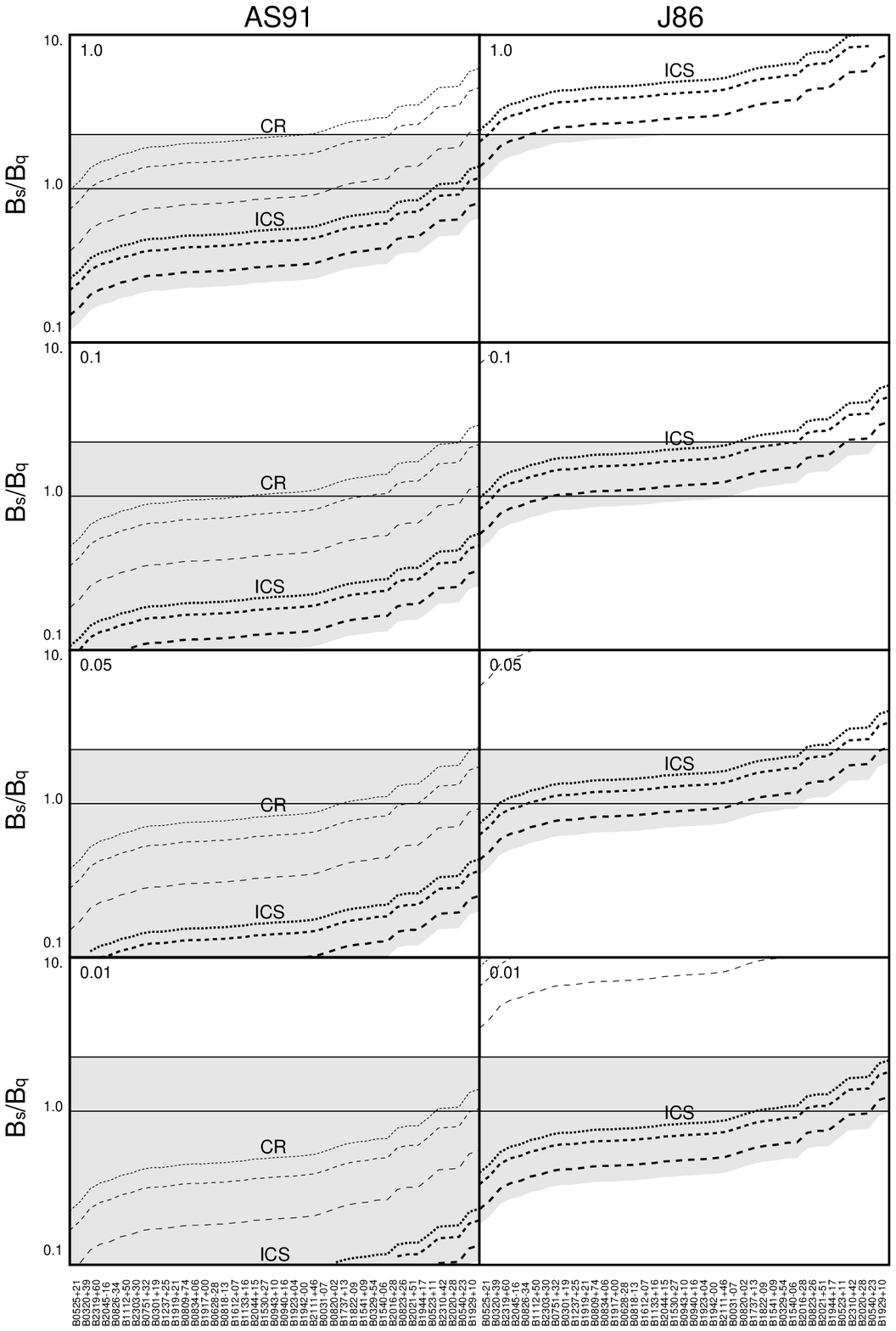}
\newpage
\plotone{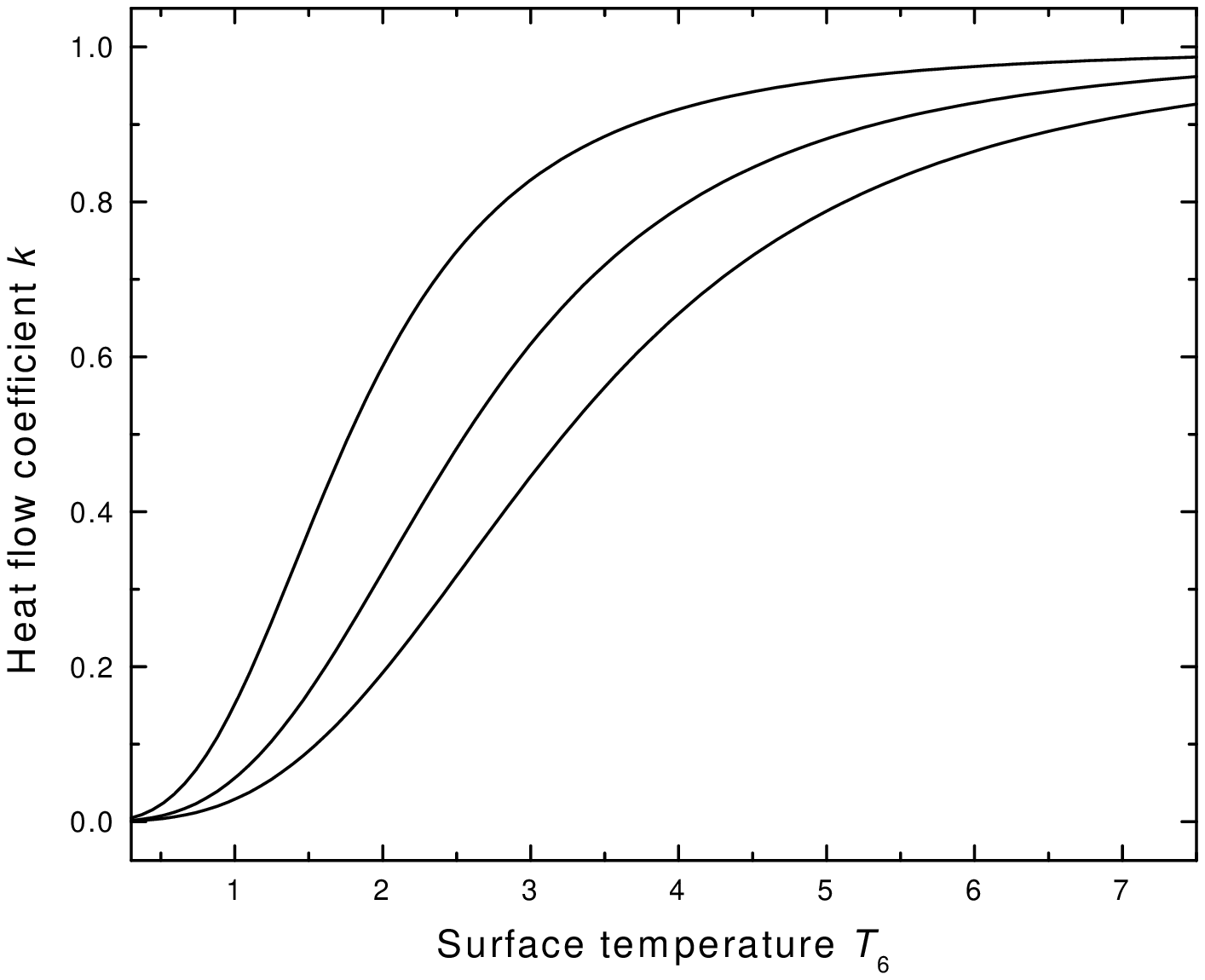}
\newpage
\plotone{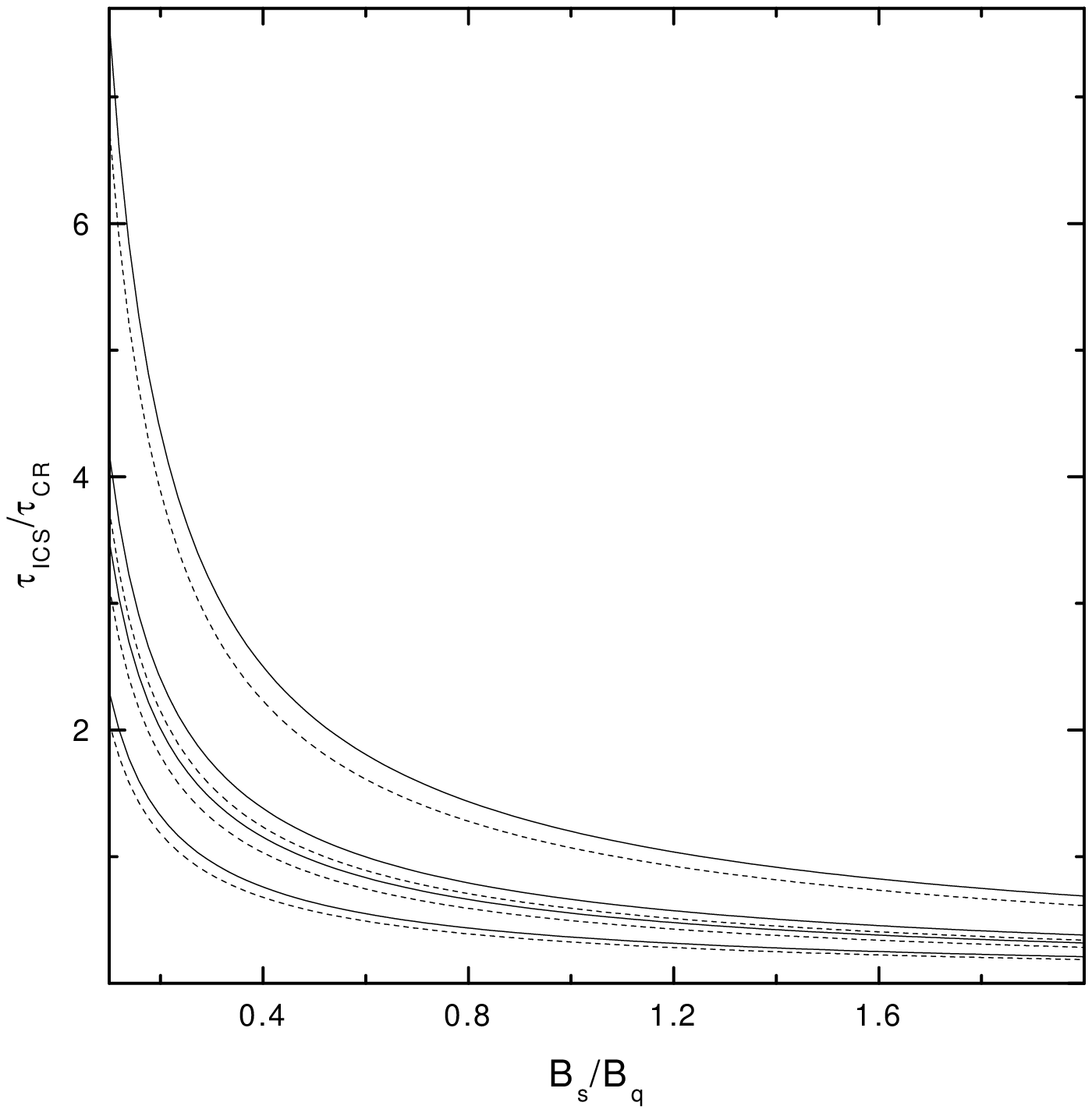}

\end{document}